\documentclass[a4paper,10pt]{article} 
\usepackage{jcappub}

\usepackage[english]{babel}
\usepackage{setspace}
\usepackage{amsfonts,amsmath,amssymb}
\usepackage{graphicx}
\usepackage{subfig}
\usepackage{booktabs}
\usepackage{tabularx}
\usepackage{caption}
\usepackage[latin1]{inputenc}
\usepackage{color}

\newcommand{\ie}{{\it i.e.}~}
\newcommand{\eg}{{\it e.g.}~}

\newcommand{\lcdm}{$\Lambda$CDM\,}
\newcommand{\ar}{\alpha_{\mathrm{Ric}}}
\newcommand{\as}{\alpha_{\mathrm{Scal}}}
\newcommand{\fid}{{\mathrm{fid}}}
\newcommand{\std}{{\mathrm{standard}}}
\newcommand{\eff}{\mathrm{eff}}

\title{Extended \lcdm: generalized non-minimal coupling for dark matter fluids}
 
\author[a,b]{Dario Bettoni,}
\author[a,b]{Stefano Liberati%
}
\author[c]{and Lorenzo Sindoni}
\affiliation[a]{SISSA/ISAS,\\
Via Bonomea 265, 34136, Trieste, Italy}
\affiliation[b]{INFN, Sezione di Trieste, \\
Via Valerio, 2, 34127, Trieste, Italy}
\affiliation[c]{Max Planck Institut f\"ur Gravitationsphysik,\\
(Albert Einstein Institute)\\
Am M\"uhlenberg 1, 14476, Golm, Germany}
\emailAdd{bettoni@sissa.it}
\emailAdd{liberati@sissa.it}
\emailAdd{sindoni@aei.mpg.de}
\abstract{In this paper we discuss a class of models that address the issue of explaining the gravitational dynamics at the galactic scale starting from a geometric point of view. Instead of claiming the existence of some hidden coupling between dark matter and baryons, or abandoning the existence of dark matter itself, we consider the possibility that dark matter and gravity have some non trivial interaction able to modify the dynamics at astrophysical scales. This interaction is implemented assuming that dark matter gets non--minimally coupled with gravity at suitably small scales and late times.  After showing the predictions of the model in the Newtonian limit we also discuss the possible origin of it non-minimal coupling. This investigation seems to suggest that phenomenological mechanisms envisaged for the dark matter dynamics, such as the Bose--Einstein condensation of dark matter halos, could be connected to this class of models.}

\date{2.8.2011}

\begin{document}

\maketitle
\flushbottom

\section{Introduction}

 Modern cosmology is living a golden age, characterized by a growing number of striking discoveries and by a stream of increasingly more accurate observations. The overall picture emerging from this wealth  of information is encoded in a simple and elegant model: the standard model of cosmology (\lcdm). 

This is just based on general relativity (GR), a cosmological constant (not theoretically understood so far), and a certain number of cosmological fluids representing the matter fields in a hydrodynamic limit description. These fluids are assumed to be minimally coupled to the gravitational field and, at least in the case of the dark matter (DM), collisionless (\ie~pressureless).

However, the quest for a better understanding of the small scale cosmological dynamics is one of the major challenges for contemporary cosmology: despite \lcdm success in explaining observations at the largest scales \cite{WMAP} (and the extremely good agreement of GR with observations at the scale of the solar system \cite{will}), observations at the galactic (and cluster) scale seem to indicate that some key ingredient is missing \cite{ferreira} from the above picture.  

On the one side there is the inability to properly fit the rotation curves of spiral galaxies \cite{rot1,rot2} and to reproduce the dark matter density profiles, the so called core--cusp problem \cite{cuco}. On the other side observations have pointed out the existence of universal relations,  the so called {\em universal conspiracy}, between baryons and dark matter, like the Tully--Fisher relation \cite{TFR} or the constant central surface density \cite{Don,Sal} and of baryons independent dark matter features, like the luminosity independent core mass for satellite galaxies of the Milky Way \cite{strig}.

If the first discrepancies might be explained through a better understanding of the baryon dynamics \cite{SN, momang}, the second kind seems more problematic as it can be  explained only at the price of inserting a fine balance between DM and baryons \cite{BTFR}.
The universal conspiracy mentioned above may represent a strong hint on the intimate nature of the dark matter and, as a consequence, a modification in the CDM paradigm might be necessary at those scales.

An alternative proposal to the DM is the well known MOND paradigm \cite{MOND,alternative}. There, in order to explain the mass discrepancy, no DM is requested. What is assumed is that there exists an acceleration scale, $a_{0}$, below which the Newtonian law is modified:
\begin{equation}
\nabla \cdot \left(\mu \left(\frac{|\nabla \phi|}{a_{0}}\right) \nabla \phi \right)=4 \pi G \rho,
\end{equation}
where $\rho$ is the baryonic mass density and $\mu(x)$ is a function whose asymptotic values are 1 for $x\gg1$ and $x$ in the opposite regime.
This formula is able to give reason of many of the observed universal properties, like the barionic Tully--Fisher relation or the constant central surface density \cite{MOND2} as they rise as intrinsic properties. Following this line of reasoning, Tensor-Vector-Scalar (TeVeS) theories of gravity or Bimetric MOND gravity have been developed as general relativistic implementations of the MOND paradigm \cite{MOND2,STVGT,Bim,fieldMOND}.

Despite its ability to reproduce some of the above mentioned universal relations (see \eg~\cite{MOND2})-- a test for which it was originally designed -- the MOND paradigm encounters a number of difficulties, especially in fitting CMB peaks \cite{MONDCMB} and in describing galaxy clusters; so much so that even in this framework one ends up requiring some amount of dark matter in order to fully explain observations at different scales \cite{MONDark,mondark2,mondark3}.

Summarizing the situation, we can say that the present understanding of the late time universe seems to be fractioned into the successes of two schemes which however {\em per se} are not able to reproduce entirely the phenomenology inferred from observations.

In an attempt to happily marry the merits of a MOND-like picture with the strengths of a CDM framework, a new framework aimed to the reconciliation of the two approaches has been suggested in \cite{RMDM}. There, it was shown how it is possible to reproduce a MONDian behavior at galactic scales in a standard CDM scenario by requiring DM  to couple with baryons in a suitable way. In other words, the MONDian behavior would emerge as an effect of the {specific} interaction between DM and baryons. If this interaction can be built so to be active at special scales and times, then one might be able to achieve the aforementioned marriage between competing models.

As said, the proposal hinges on a realization of an effective non--minimal coupling between dark matter and the gravitational field, which is otherwise described as in GR. The origin of this coupling should not be viewed necessarily as a form of fundamental interaction between DM and baryons, but rather as a geometric effect due to a non trivial interaction between DM and gravity. 

This proposal has (at least) two distinctive features: one is that DM fluid is no longer perfect at the scales at which the non--minimal coupling is relevant; the other is that the physical metric (the one along which baryons move) gets redefined in a way that depends only on how DM and gravity are coupled. This leads to a modified effective dynamics for the fluid {\it and} for the gravitational field even in the Newtonian limit, as we shall see. The first feature may address the problems of dark matter density in halos, while the second one may provide an explanation to the above mentioned \emph{universal conspiracy}.

Here we will further extend the analysis of this coupling between DM and gravity. We will consider a dark matter fluid in a standard General Relativistic scenario in which we have dropped the assumption of minimal coupling and study the consequences of the most generic coupling that can be built keeping second order gravity. The possibility that DM has non trivial geometric properties has not yet received full consideration, and a detailed study of this topic might shed light not only on the phenomenology of small scales dynamics but also on possible interplays between matter fields and gravity.\\

The plan of the paper is as follows.
In section \ref{sec2} we will introduce the concept of geometric matter action, briefly revising past approaches to the subject, while in section \ref{sec3} we will extend this idea to a non--minimally coupled  fluid while in section \ref{sec4} we will describe the phenomenology expected from our model when applied to cosmological fluids. In section \ref{sec5} we shall then discuss some ideas about the origin of the non--minimal coupling before to finally draw our conclusions in section \ref{sec6}.

\section{Geometric action for dark matter}
\label{sec2}

In general relativity, matter minimally couples to a metric, the so called \emph{physical metric}, and this metric is assumed to be the same that describes the dynamics of the gravitational field, the \emph{gravitational metric}. However there is no first principle that forces to do this choice, if not the requirement of simplicity\footnote{Indeed, if we do not introduce additional fields and we want to keep the equations of motion local, the option that we are considering in the following is the only one remaining.}. Relaxing the assumption allows for the physical metric and the gravitational to not coincide, as the physical metric can be made out of a combination of gravitational metric and extra field, the most common example being conformal relations.

It was shown in \cite{disf} that the most generic transformation between physical and gravitational metric that preserves causality and the weak equivalence principle (WEP) can be of the general form:
\begin{equation}
\widetilde{g}_{\mu \nu} = A(\phi)g_{\mu \nu} + B(\phi)\nabla_{\mu}\phi \nabla_{\nu}\phi, \label{disf}
\end{equation}
where $\phi$ is an extra scalar field while $A$ and $B$ are functions on which causality, \ie the existence of light cones, imposes constraints.

Relation (\ref{disf}) is the so called disformal transformation which finds several applications in the comsological contest, \eg in dark energy models \cite{mota} or in inflationary models \cite{disfinfl}.  Notice that here we work in an effective hydrodynamical framework. This means that many of the constraints that must be satisfied by effective quantum field theories \cite{adams} are not immediately translated in our framework.

From the study of scalar--tensor theories of gravity we know that if we express both the gravitational action and the matter action in terms of the physical metric (Jordan frame), then extra field gets non--minimally coupled to gravity. On the contrary if one wants to express everything in terms of the gravitational metric (Einstein frame), the standard Einstein--Hilbert action is recovered but matter degrees of freedom get coupled with the extra scalar field \cite{far}. 

The introduction of an \emph{ad hoc} scalar field to solve the mass discrepancy puzzle may seem not physically justified. If one identifies the extra scalar field with the DM field and takes into account the presence of baryons, then it is natural to ask whether and how the coupling can modify the dynamics of baryons.

The basic idea of \cite{RMDM} is to extend this picture, keeping the geometrical interpretation of the interaction term between dark matter and baryonic fluids, but including the possibility of disformal transformations. In this way, the coupling between DM and baryons can be designed to modify the dynamics of baryons, to mimic as close as possible the MONDian phenomenology at galactic scales, keeping the model within the CDM paradigm.

The starting point is the following action:
\begin{equation}
S = S_{EH}[g_{\mu \nu}] + S_{Mat}[g_{\mu \nu},\psi]+S_{DM}[g_{\mu \nu},\phi]+S_{Int}[g_{\mu \nu}, \psi,\phi],
\end{equation}
made by the Einstein--Hilbert action,\footnote{We do not consider here the possible presence of a cosmological constant
given that we aim at discussing just the modified dynamics of
non-minimally coupled dark matter. However, once phenomenological
signatures of the framework will have to be discussed, it might be
possible that the so called "dark degeneracy"\cite{kunz,cota} will have to be
addressed.} the standard baryonic and dark matter actions, and the interaction one respectively. Here, the (scalar) fields $\psi$ and $\phi$ are thought as collective variables encoding baryons and DM particles, respectively. In order to reproduce a MONDian behavior the coupling should be able to produce an extra--force that applies to baryons only. This can be achieved assuming that the interaction term is such that
\begin{equation}
S_{Mat}[\psi, g_{\mu \nu}] + S_{Int}[g_{\mu \nu}, \psi,\phi] \approx S_{Mat}[\psi,g_{\mu \nu}+h_{\mu \nu}],
\end{equation}
where the net effect of the interaction can be then summarized into a modification of the metric felt by the baryons. Up to order $\mathcal{O}(h^{2})$ the interaction term is:
\begin{equation}
S_{Int}[g_{\mu \nu}, \psi,\phi] = \frac{1}{2c}\int d^{4}x\,\sqrt{g}\,T^{\mu \nu}_{Mat}h_{\mu \nu},
\end{equation}
where the detailed shape of $h_{\mu \nu}$ is fixed in such a way to give MONDian behavior at the scales of interests.\\
Up to order $\mathcal{O}(h^{2})$ the theory is bimetric and if now we express the full actions in terms of the physical metric $\widetilde{g}_{\mu \nu} \equiv g_{\mu \nu} + h_{\mu \nu}$ we get that the DM field gets non--minimally coupled to gravity in the following way
\begin{equation}
S = S_{EH}[\widetilde{g}] + S_{Mat}[\widetilde{g},\psi]+S_{DM}[\widetilde{g},\widetilde{\chi}] 
-\frac{1}{16 \pi G} \int d^{4}x\,\sqrt{-g}(G^{\mu \nu}(\widetilde{g}) + \Lambda \widetilde{g}^{\mu \nu})h_{\mu \nu},
\end{equation}
where $G^{\mu \nu}$ is the Einstein tensor expressed in terms of the physical metric and where the DM field has been implicitly redefined $\chi \rightarrow \widetilde{\chi}$.
It is this geometric interaction, with a suitable choice of the tensor $h_{\mu\nu}$, that allows to reproduce the MONDian regime.

This approach has some unpleasant ambiguities and interpretational shortcomings. First of all, the model has been formulated in terms of scalar fields, which are indeed relevant to understand the behavior of certain gravitational phenomena, but are not, perhaps, the best tools to describe cosmological fluids, especially at the galactic level. Second, they were introduced as phenomenological fields, to be used as fluid variables, without any reference to the underlying particle physics. We
will not address the derivation of these fields and their dynamics from
a specific model, although derivations of hydrodynamic fields and their
couplings from a fundamental level can be envisaged, see e.g. \cite{khoury,chameleon,conlon,wett}. Third, the needed interaction term must be carefully designed, at the field--theoretic level, in a way that seems very counterintuitive. Finally, the very nature of the dark matter field $\phi$ seems ambiguous: in such a framework, is it a fluid field, or an additional gravitational potential (\ie a fifth force)?

To address these issues, we leave the field--theoretic language and use, in a systematic way, the relativistic fluid dynamics language, which is certainly more reasonable to apply for describing gravitational phenomena of galaxies and larger systems,
provided that the averaging scale $L$ necessary to define the fluid variables is small enough, compared to the size of the structures that we want to analyze.

Furthermore we are entitled to use such a description for dark matter since all its quantum properties are negligible or averaged out at the scales of interest.\footnote{An exception to this statement could be realized in the presence of torsion or in BEC, as we will discuss later.}

The case of non-minimally coupled fluids was first considered in \cite{goen}, where some basic rules for building a consistent theory of gravity with a non--minimal coupling (NMC) matter sector were posed. Due to the particular coupling to gravity, in general a NMC perfect fluid is no longer perfect, being equipped with anisotropic stresses and momentum flows that are originated by the coupling to the curvature tensors. Therefore, it is reasonable to expect that all this corrections may play a role in galactic scale cosmology, even if, from a naive point of view, one might expect that in regions of very small curvature these terms should be negligible. In the next section we will then reconsider this analysis, extending it to a broader class of non--minimal couplings and investigate to which sort of bimetric theories this leads to. 

\section{Non--minimally coupled dark matter fluid}
\label{sec3}

Consider a system in which DM is described as a perfect fluid with a barotropic equation of state which couples non minimally to gravity. The easiest way to do this is to couple a scalar function of the DM variables to the Ricci scalar, adding to the Lagrangian a term like $f(R)\mathcal{L}_{dm}$, where $f(R)$ is a generic function of the Ricci scalar and $\mathcal{L}_{dm}$ would be the DM Lagrangian (for a thorough discussion of these models see, for instance, \eg \cite{LmR} (but see also \cite{bertolami2,bertolami3} for further investigations on the class of models). However this particular scalar coupling would not affect the propagation of light rays, given that the Maxwell action is conformally invariant. Such a coupling would not be enough for enhancing gravitational lensing as it seems necessary in order to account for the observed DM phenomenology \cite{weaklens,weaklens2,weaklens3}. Obviously a more general coupling is required. 

Moreover, since the deviations from \lcdm are effective only at galactic scales (where densities are higher than the cosmological ones), the coupling has to be  active only above a certain density threshold, at late times. This would fix a minimum scale at which deviations from \lcdm are expected to be present. 
Clearly, we need to elaborate more on this, since we do not want to spoil the description of equally high density but spatially homogeneous early universe cosmology.
We will come back to this point in the concluding section.

It is rather clear that, working with perfect fluids, there are not many possibilities to couple DM  to gravity, given that we have at our disposal only scalars and the four vector field encoding the four velocity at each spacetime point. 
If we add the constraint that we want to keep the gravitational field equations of second order in the metric tensor, so that we
do not introduce additional physical modes for the graviton\footnote{These extra modes would be dangerous for our program, given that they could affect gravitational physics in regimes where we do not want to have {noticeable discrepancies with the standard predictions of GR}.}, there are only two possible couplings, namely
\begin{equation}
R\psi(\rho) \qquad \mathrm{and} \qquad R_{\mu\nu} \xi(\rho) u^{\mu}u^{\nu}.
\end{equation}
The requirement of second order gravitational equations forces us to consider terms which contain the second order derivatives of the metric at most linearly, \ie terms that are linear in the Riemann tensor. The latter must be contracted with suitable tensorial structures in order to produce scalars that can enter an action. We have only the metric tensor and the vector $u^{\mu}$. Therefore, we end up with only the two terms that we have just introduced, as long as curvature is concerned. 
Furthermore, if we use perfect barotropic fluid, the residual information about the coupling can be parametrized completely with two arbitrary functions of the mass density (see again \cite{chameleon} for another example of density dependent couplings ).

Before moving on, let us remind the reader that we are denoting as $\rho$ the mass density, that $u_{\mu}u^{\mu}=-1$ and that for the rest we are following the treatment presented in \cite{PFA}.

\subsection{Action and equations of motion}
As it has been discussed at length, we are considering a case in which the baryons, by definition, are coupled minimally to the
physical metric, while the dark matter fluid is coupled to it non-minimally. As a consequence, the equation of motion for baryons (continuity and Euler) are the textbook ones. Therefore, in this section, we will consider only the case of a non--minimally coupled fluid. The action will be,
\begin{equation}
S = \frac{c^{3}}{16 \pi G_N}\int d^{4}x\,\sqrt{-g} \Psi(\rho)R+ \frac{\ar c^{3}}{16 \pi G_N}\int d^{4}x\,\sqrt{-g} R_{\mu \nu}\xi(\rho)u^{\mu}u^{\nu} + S_{DM}\label{action}.
\end{equation}
The DM fluid action is the action for  perfect fluid as reported in \cite{PFA}:
\begin{equation}
S_{DM} =  - 2c\int d^{4}x \, \left[ \sqrt{-g} \rho(n,s) + J^{\mu}(\phi_{,\mu} + s \theta_{,\mu}+ \beta_{A}\alpha^{A}_{,\mu}) \right],
\end{equation}
where $n$ is the particle number density, $s$ is the entropy per particle and the second term implement the constraints for the flow of perfect fluid. The densitized four vector $J^{\mu}$ is related to the fluid variables:
\begin{equation}
J^{\mu} = n u^{\mu} \sqrt{-g}\,.
\end{equation}
The function $\Psi(\rho) = 1 + \as\psi(\rho)$ controls the coupling of the dark fluid to the Ricci scalar, while the function $\xi(\rho)$ mediates the coupling to the Ricci tensor. Both these functions are dimensionless, and hence they must involve, for dimensional reasons, at least another density parameter $\rho_{*}$ which sets a characteristic, phenomenological, scale of the model. Finally, the dimensionless constants $\as,\ar$ control the strength of the non-minimal coupling. In fact, they could be reabsorbed in the functions $\psi$ and $\xi$, without loss of generality. However, it is useful to keep them explicit since they can be used as dimensionless parameters for an expansion whenever the non--minimal coupling is expected to be a subdominant effect.

Following the standard rules of the calculus of variations, one can find the equations of motion for the system. The equations obtained varying the action with respect to the metric are:
\begin{equation*}
G^{\mu\nu} = \frac{8\pi G_{N}}{c^{2}}\rho u^{\mu}u^{\nu} + \alpha_{scal} \left( -\psi(\rho) G^{\mu\nu} - \Box \psi g^{\mu\nu} + \nabla^{\mu}\nabla^{\nu} \psi - \frac{R}{2} \psi' \rho H^{\mu\nu}  \right)+\end{equation*}
\begin{equation}+\frac{\alpha_{Ric}}{2} \left( - \Box t^{\mu\nu} + \nabla_{\rho} \nabla^{\mu} t^{\rho \nu} + \nabla_{\rho} \nabla^{\nu} t^{\rho \mu} -g^{\mu\nu} \nabla_{\alpha} \nabla_{\beta} t^{\alpha \beta} +
R_{\alpha\beta}u^{\alpha}u^{\beta} \left( \xi - \frac{1}{2} \xi' \rho \right)H^{\mu\nu}\right) ,
\label{eq:einstein}
\end{equation}
where we have introduced the notation
\begin{equation}
H^{\mu\nu} = g^{\mu\nu}+u^{\mu}u^{\nu},
\end{equation}
for the projector on the subspace orthogonal to $u^{\mu}$, and the tensor
\begin{equation}
t^{\alpha\beta} = \xi(\rho) u^{\alpha} u^{\beta},
\end{equation}
to slightly simplify the expressions.

As one can easily see, the Einstein equation do not contains higher derivatives of the metric tensor. However, in addition to the stress energy tensor (SET) for a fluid made of dust, there are a certain number of terms that concur to define an effective SET, depending on higher derivatives of the fluid variables and on the curvature.

This is indeed what should be expected, given that the basic idea of non--minimal coupling is that the field, or fluid, is able to probe geometry on a given length scale, not only point--wise as in the standard case.

The equations of motion for the fluid can be obtained either varying the action with respect to the various fluid fields, or by using the Bianchi identities on the modified Einstein equations. We will discuss them elsewhere, in full generality. For the purpose of this paper we shall limit ourselves to their Newtonian limit, which is the relevant regime to discuss galactic dynamics.

\subsection{Newtonian limit}
To properly discuss the Newtonian limit, it is important to carefully work out the weak field limit of Eq.\eqref{eq:einstein}.
Indeed, it is important to understand how to establish a comparison between terms that have different physical dimensions.
As we have discussed, $\psi,\xi$ are dimensionless, as $\as,\ar$. Assume that the typical size of the system under consideration is of order $\ell$. With this length scale, we can redefine the coordinates in such a way that they are dimensionless:
\begin{equation}
x^{\mu} \rightarrow y^{\mu} = x^{\mu}/\ell.
\end{equation}
This allows to rewrite the Einstein equations in such a way that all the terms are dimensionless, and hence enables us to compare the different terms in a consistent way. The weak field limit is achieved whenever the curvature radius is much larger than the size of the system, \ie
\begin{equation}
|G_{\mu\nu}| \ll \ell^{-2}.
\end{equation}
For consistency, then, $\xi,\psi$ must be small quantities, as well as the properly normalized energy density.
In short, the weak field limit is achieved when the following condition holds on the SET:
\begin{equation}
\psi \simeq \xi \simeq \frac{8\pi G_{N}}{c^{2}} \rho \ell^{2} \ll1.
\end{equation}

To keep this systematically under control, we need to reparametrize $\psi$ and $\xi$ in order to
introduce $c^{2}$ factors that will be used to take into account that $\psi,\xi$ are as small as $\rho$, even if they are measured in different units. We can always rescale $\psi$ and $\xi$ so that

\begin{equation}
\psi(\rho) = \frac{8\pi G_{N}}{c^{2}} \ell^{2} \rho_{\fid} \tilde{\psi}(\rho),
\end{equation}
\begin{equation}
\xi(\rho) = \frac{8\pi G_{N}}{c^{2}} \ell^{2} \rho_{\fid} \tilde{\xi}(\rho),
\end{equation}
where $\rho_{\fid}$ is a fiducial density that has to be determined by some condition, like
\begin{equation}
\psi(\rho_{\fid}) = \frac{8\pi G_{N}}{c^{2}} \ell^{2} \rho_{\fid} \tilde{\psi}(\rho_{\fid})= \frac{8\pi G_{N}}{c^{2}}\rho_{\fid} \ell^{2},
\end{equation}
that is
\begin{equation}
 \tilde{\psi}(\rho_{\fid})=1.
\end{equation}
With this treatment, we can consistently extract the weak field limit of the equations Eq. \eqref{eq:einstein}, taking systematically into account the relative size of the various terms. 
If we define 
\begin{equation}
g_{\mu\nu} = \eta_{\mu\nu} + \gamma_{\mu\nu} ; \qquad \bar{\gamma}_{\mu\nu} = \gamma_{\mu\nu} - \frac{1}{2} \eta_{\mu\nu} \gamma; \qquad \gamma = \eta^{\mu\nu} \gamma_{\mu\nu},
\end{equation}
the modified Einstein equation, in the weak field limit, in the transverse gauge, reads:
\begin{equation}
-\frac{1}{2} \Box \bar{\gamma}_{\mu\nu} = \frac{8\pi G_{N}}{c^{2}}  \left\{- \rho_{fid} \left[ 
\as \left(\eta_{\mu\nu} \Box \tilde{\psi} - \partial_\mu\partial_\nu \tilde{\psi} \right) + \frac{\ar}{2} \Omega_{\mu\nu}
\right] + \rho u^{\mu}u^{\nu}\right\},\end{equation}
where
\begin{equation}
\Omega_{\mu\nu} =
\delta_{\mu}^{0}\delta_{\nu}^{0}\Box \tilde{\xi}(\rho) 
- \delta_{\nu}^{0} \partial_{0} \partial_{\mu}\tilde{\xi}(\rho)
- \delta_{\mu}^{0} \partial_{0} \partial_{\nu}\tilde{\xi}(\rho)
+\eta_{\mu\nu} \partial_{0}\partial_{0}\tilde{\xi}(\rho). 
\end{equation}

As one immediately sees, the effect of the non--minimal coupling is still present, even in the weak field limit, and the fluid is not behaving as a perfect fluid in Minkowski spacetime: the non--minimal coupling has generated a SET which contains additional terms,
constructed out of the derivatives of the fluid variables. These terms will enter the fluid equations, that we will briefly discuss in the appendix. For now, we will simply point out their presence.

Putting everything together, and considering the static, nonrelativistic limit (\ie the $c^{2} \rightarrow \infty$ limit), we get the Poisson equation for the Newtonian gravitational field:
\begin{equation}
\nabla^{2} \Phi_{N} = 4 \pi G_{N} \left( \rho -\frac{ \ar}{2} \rho_{\fid} \nabla^2 \tilde{\xi} + \as \rho_{\fid} \nabla^2 \tilde{\psi} \right).
\end{equation}
As it happens for the fluid equations, even the Newtonian potential has as sources not only the mass density $\rho$, but also a certain number of derivative terms. In light of the fact that the operative definition of the dark matter mass density is given in terms of the density that enters the right hand side of the equation of motion for the gravitational field (either the Poisson equation or the general-relativistic version), one might define an effective mass density and effective stresses, that do not coincide with those that are defined out of the fluid action when the non--minimal coupling is absent.

\section{Phenomenological constraints}
\label{sec4}

Now that we have lied down our model and analyzed its Newtonian limit, we are in the condition to discuss more accurately its predictions in different regimes and consequently use current observations to bound it (in particular by constraining the behavior of the functions $\psi(\rho),\xi(\rho)$). There are two obvious such regimes at which the model has to offer new phenomenology: galactic dynamics and cosmology. However, to be viable, any modified gravity model (in a broad sense) must be compatible with solar system constraints on gravitational phenomena. We shall hence start our discussion from here.

\paragraph{Solar System scales:} Of course, our model must reduce to general relativity (or be very close to it) at these scales.  In particular, if we impose that $\psi(0)=\xi(0)=0$, we are sure that the dynamics of a purely baryonic system will be described by general relativity without corrections. Given that, at the level of solar system, it is safe to say that the density of baryonic matter
is much larger that the density of dark matter, this condition ensures that the agreement with observational constraints will be achieved, provided that $\as,\ar$ are not too large.

\paragraph{Galactic dynamics:} We have shown that the Poisson equation gets modified by a term which depends on gradients of the density.  This means that the more inhomogeneous a distribution of DM is, the stronger is the effect. As a consequence, structures may grow faster or slower than expected, according to the structure of the additional terms, and, ultimately, to the signs of the coupling constants $\as,\ar$.

As we have mentioned, the NMC coupling also generates a pressure term, which is structured in two components. 
On the one hand, there is an isotropic pressure that again is related to gradients of the density. This is a key feature as pressure may stabilize halo's cores preventing the formation of cusps, given that its magnitude increases as the inhomogeneity increases. 

On the other hand, there is an anisotropic pressure term which represent a distinguishing feature of our model. In standard CDM model particles forming halos are collisionless and hence they have no global collective motion. This anisotropic pressure may generate a net overall rotation of DM halos which modifies the caustic structure of the infalling dark matter particles with respect to the irrotational flow. There is convincing evidence that such overall rotation can lead to a caustic structure closer to the observed one \cite{caustics}.

These generic features of the here proposed models are definitely interesting, given that they are able to affect, in a rather transparent way, some important issues of the galactic dynamics, which are not easy to address within standard
\lcdm approaches. 

Given that the puzzles related to mass discrepancies are harder to address at the galactic scale, one needs the NMC terms to be larger in these regimes and consequently to set $\rho_{*} \approx \rho_{gal}$;  basically assuming that the functions $\psi,\xi$ will attain their maxima in this density regime.

\paragraph{Cosmology:}
As pointed out earlier, a key feature of this model is the presence of spatial gradients of the density in the non-relativistic limit. However, in the full relativistic theory, not only spatial derivatives, but also time derivatives are relevant, and the additional terms
might be active even in spatially homogeneous cases. The NMC may affect the cosmological evolution in a dramatic way that might lead to a sharp contrast with the observations, whenever the time derivatives become relevant, \ie at sufficiently early times in cosmology.

Consider the  FFRW metric:
\begin{equation}
ds^{2} = e^{2n(t)}dt^{2} + e^{2a(t)}d\mathbf{x}^{2}.
\end{equation}
We can compute the Lagrangian for our model inserting the metric defined by the above line element in the action \ref{action}. This gives the following effective Lagrangian density (where a boundary term has been discarded):
\begin{equation}
\mathcal{L}_{grav} = \frac{e^{-n + 3a}}{16 \pi G}\left\{-6\left[ \dot{a}^{2}+(\as \psi' + \ar \xi')\dot{a}\dot{\rho}\right] - 6\as \psi \dot{a}^{2}-3\ar \xi \dot{a}^{2}\right\},
\end{equation}
to which the fluid Lagrangian density has to be added.

We now want to recover, at large scales and at early times, the \lcdm model. Given that on large scales we can safely use spatially homogeneous configurations, we need only to take care of temporal gradients. 
To be sure that these are not effective in changing much the dynamics away from \lcdm, we need to ask that the non-minimal coupling terms disappear for sufficiently dense or hot fluid. 

This requirement suggests that our functions $\psi,\xi$ must be strongly peaked around $\rho_{*}$.
Concretely, this means that as the density reaches the value $\rho_{*}$, then we get modified cosmological evolution, until $\rho$ drops well below $\rho_{*}$. If we take today cosmological DM density to be of the order of $.24 \times 10^{-29}g/cm^{3}$ and the reference density to be $\rho_{*} \approx 10^{-21}g/cm^{3}$ -- the typical value for dwarf spheroidal galaxies -- we get that:
\begin{equation}
1+z_{*} = \Big(\frac{\rho_{*}}{\rho^{dm}_{0}}\Big)^{1/3} \sim 700.
\end{equation}
This seems to indicate that our model may strongly affect the background evolution in a small redshift window in the matter dominated era, something for which there is no evidence. Nonetheless, it is not obvious that these modifications of the early
universe dynamics could not be made compatible with current observations. We just notice here that the latter are normally able to cast strong constraints, for example via the CMB or Big Bang Nucleosynthesis.

However, the above discussion holds only if the NMC is taken as fundamental so that its action is present all along the whole history of the universe. We have no reason to believe that this is true and we shall argue below reasons to expect the contrary.

\section{Origin of the non-minimal coupling}
\label{sec5}

Up to now we have not given any reason why only dark matter should couple non minimally to gravity. Furthermore, we have seen that a parametrization of the functions $\xi,\psi$ with only densities might lead to discrepancies from the expected behavior starting at relatively large redshift. Therefore, to address this tension we need to understand more of the possible mechanisms that can lead to the non-minimal coupling as a phenomenologically more accurate description of the dark matter fluid.

The fact that only DM couples to gravity in a non trivial way may be seen as a violation of the weak equivalence principle (WEP). However, here we are dealing with fluids, not elementary particles. Hence WEP is safe as long as single particles have the same coupling with gravity, while the WEP can be nonetheless violated at the level of the collective behavior of the fluid.

There are two main mechanisms that may can produce a non--minimal coupling: either it appears through an averaging procedure that brings from particles to fluids or it can emergence from some collective behavior of the DM particles.

In the first case there is a scale, the averaging scale (which depends on the number density of the DM particles, on statistical grounds). If these are heavy, the size of the averaging scale may be large enough to be comparable with the curvature radius of the galaxy and hence generate a non-minimal coupling, given that the minimal cell needed to define a fluid element is able to probe geometry in a nonlocal way, becoming explicitly sensitive to curvature. In this case, however, the reasoning applies to DM as well as to baryons, for which the non-minimal coupling does not seem so well motivated (see, however \cite{bertolami2,bertolami3,harkorot} for a proposal to explain dark matter as an effect of non--minimally coupled baryons).

The second picture is related to the possibility for DM particles to develop a macroscopic coherence length.  One recently explored instance of this is Bose--Einstein condensation (BEC) \cite{harkoBEC, chavanis}.\footnote{See also \cite{hu} for a slightly different approach to the solution of the core-cusp problem, but also \cite{woo} for a counterexample to it.} In BEC, the condensate possesses a characteristic
coherence length, the healing length, that controls the deviation of the fluid dynamics of the condensate from the one of
an ordinary perfect fluid. The BEC option seems to be rather intriguing for our model, given that it would be able to
reconcile the puzzle between the large density MONDian behavior of galaxies and the large density \lcdm behavior of early universe. Notice that, while BEC is a macroscopic quantum configuration of matter,
it admits a rather standard hydrodynamical description, given by the
Gross-Pitaevski equation for the classical condensate wave function.

The answer to the puzzle would be that the functions $\psi,\xi$, besides the density, depend on the \emph{temperature} of the fluid itself: if the temperature of the fluid is smaller than the critical temperature, condensation sets in, and with it non-minimal coupling (provided that the coherence length is large enough). On the contrary, if the fluid is too hot, the condensation is impossible, and the fluid behaves like an ordinary fluids.
Noticeably, in trapped BECs, the critical temperature increases with the depth of the potential well in which they are confined. Similarly, clumping of dark matter halos at the galactic scale might raise the critical temperature so that it is actually larger than
the temperature of the dark matter fluid, triggering condensation. On the contrary, large density but too high temperature, as in high redshift universe, might make condensation impossible.

\section{Conclusions}
\label{sec6}

In this paper we investigated how a non-minimally coupled dark matter fluid can modify the dynamics in the nonrelativistic regime with respect to the standard CDM picture. Let us summarize the key points.

\begin{itemize}
\item
We have found that such coupling is able to modify the Poisson equation, introducing an additional source term for the gravitational field. This implies that the source for the gravitational field is not just the number density: also inhomogeneities in the distributions of particles affect the gravitational field. 
\item
The non--minimally coupled fluid is described by a rich stress tensor. The pressure term has two components: isotropic and anisotropic. While at a first glance they might seem unpleasant, they can lead to two welcome effects on DM halos. The isotropic part can stabilize DM distribution and avoid the formation of cusps leading to a more cored density profile, while the anisotropic can give a preferred direction leading to an overall rotation of the DM distribution.
\item
With this model we may be able to address some of the problems that \lcdm is suffering, by reproducing, at suitable scales, a MOND--like behavior. We have seen that gravitational dynamics becomes nontrivial, and might ultimately lead to the appearance of a full fledged MOND regime. However, we do not have yet established a one to one correspondence between our model and MOND in its
traditional incarnation. Actually, this correspondence could be achieved only if baryons would end up tracking DM. Indeed, if we interpret the extra contribution emerging from the modified dynamics of MOND as DM, it would be nonetheless locked to the baryon density. To settle this point, the detailed analysis of the gravitational dynamics of a galaxy, within this model, is required. While the form of the Poisson equation gives the feeling that at least a slight tracking will be present, it worth stressing that this model will generically show a richer phenomenology than MOND and could at most mimic it in some regimes.
\end{itemize}

Non--minimal coupling is not a new topic in gravitational physics. However, as we have discussed in this paper, there is today a convergence of evidences and ideas that are pointing in the direction of effective descriptions of cosmological fluids mainly at galactic scales, that include additional, phenomenological quantities, and seem able to explain current observations. Nonetheless, the physical mechanisms that would introduce such quantities and justify a departure from \lcdm are largely unknown, and probably will be ultimately clarified only when the exact properties of the dark matter particles will be understood.

In this sense, it is intriguing the idea that, due to the formation of deep enough gravitational potential wells, a dark matter condensation can be triggered at suitable scales and times and that this phenomenon might be indeed considered as a candidate for the physical origin of the here generalized non-minimal coupling. While this is an exciting perspective worth exploring, we feel that some caution should be used, especially when applying our laboratory based intuition of BEC features to cosmology.

First of all, for this mechanism to take place and be effective in cosmology, a tight balance between the microscopic properties of the dark matter bosons and the various macroscopic parameters observed must be realized (\eg the required size of the healing length, needed to solve the cusp problem, appears to be of the order of some parsecs). 

Secondly, there is a big qualitative difference between the fluid dynamics of a standard BEC and the fluid dynamics of the NMC fluid that we have explored in this paper. In fact, the pressure of the BEC gets corrected by the so--called quantum potential, 
\begin{equation}
p_{BEC} = p_{\mathrm{hydro}}(\rho) + V_{Q}; \qquad V_{Q}= -\frac{\hbar^{2}}{2m} \frac{\nabla^{2}\rho^{1/2}}{\rho^{1/2}}.
\end{equation}
This gives rise to a dependence of the pressure of the fluid on the gradients of the density closely resembling what found in the Newtonian limit of our model.  However, it is easy to see that no anisotropic stresses are present in this case while the NMC seems to lead generically to the appearance of off-diagonal terms in the SET. This issue probably requires a more accurate analysis possibly by considering more general theoretical settings for the condensation with respect to the standard one based on scalar fields.\footnote{Furthermore, notice that the quantum pressure is not just the Laplacian of a function of the density, while our equations tell us that the additional pressure term (for the non-minimally coupled fluid in the Newtonian limit) will be always a Laplacian of a function of the density.} We live this to future explorations.

In conclusion, we have here lied down a new framework for dark matter fluid dynamics which seems able to conciliate several ideas which have been advanced to improve on \lcdm predictions. We think that now further investigations can, on the one side, focus on extracting more detailed predictions from the model, for example by considering the issue of structure formation. On the other side, is worth exploring the above mentioned issue about the origin of the extended non--minimal coupling of dark matter both for its connection with extant ideas about the nature of dark matter (BEC) as well as for its implications with regard the particle physics nature of this evasive cosmological component.
We hope to be able to develop both these lines of research in the next future.

\acknowledgments{The authors wish to thank Paolo Salucci and Valeria Pettorino for illuminating discussions.}

\appendix
\section{Fluid equations and their stability}

The shape of the Einstein's equations suggests that we identify an effective SET of the form:
\begin{multline*}
T_{\mu\nu}^\eff  = T_{\mu\nu}^{\std} + \alpha_{scal} \ell^2 \rho_\fid \left( -\psi(\rho) G_{\mu\nu} - \Box \psi g_{\mu\nu} + \nabla_{\mu}\nabla_{\nu} \psi - \frac{R}{2} \psi' \rho H_{\mu\nu}  \right)+\\
+\frac{\alpha_{Ric}}{2} \ell^2 \rho_\fid\left( - \Box t_{\mu\nu} + \nabla_{\rho} \nabla_{\mu} t^{\rho}_{ \nu} + \nabla_{\rho} \nabla_{\nu} t^{\rho}_{ \mu} -g_{\mu\nu} \nabla_{\alpha} \nabla_{\beta} t^{\alpha \beta} +
R_{\alpha\beta}u^{\alpha}u^{\beta} \left( \xi - \frac{1}{2} \xi' \rho \right)H_{\mu\nu}\right)  ,
\label{eq:SETeff}
\end{multline*}

where we have removed the tilde from the functions $\psi,\xi$. As we have said, $\ell$ represent the linear scale of the system we are going to describe, while $\rho_\fid$ is a suitable normalization mass density. 

The fluid equations of motion can be derived from the Bianchi identities. Indeed, from the Einstein equations for the system:
\begin{equation}
\nabla^{\mu}T_{\mu\nu}^\eff=0.
\end{equation}
In order to compute the covariant derivative of the SET notice  that:
\begin{equation}
\nabla^{\mu}H_{\mu\nu} = \vartheta u_\nu + u^{\mu}\nabla_{\mu} u_{\nu},
\end{equation}
where
\begin{equation}
\vartheta = \nabla_{\mu} u^{\mu} ,
\end{equation}
is not yet defined as the expansion of the bundle of geodesics (but rather as the expansion of the bundle of curves whose tangent vector field is $u^{\mu}$).

Then using the commutation rules for the covariant derivatives:
\begin{equation}
\nabla_{\alpha} \nabla_{\beta} \nabla_{\nu} f = \nabla_{\alpha} \nabla_{\nu} \nabla_{\beta} f = \nabla_{\nu} \nabla_{\alpha} \nabla_{\beta} f  - R^{\rho}_{\beta \alpha \nu} \nabla_{\rho} f,
\end{equation}
and
\begin{equation}
 \nabla_{\rho} \nabla_{\nu} t^{\rho \mu} = \nabla_{\nu} \nabla_{\rho} t^{\rho \mu} 
 + R^{\rho}_{\sigma \rho \nu} t^{\sigma \mu} +R^{\mu}_{\sigma \rho \nu} t^{\sigma \rho} ,
\end{equation}
 we have that the complete fluid equations reduce to the following expression:
 \begin{equation*}
\nabla^{\mu} T_{\mu\nu}^{\std} = -\as \ell^2 \rho_{\fid} \left( \frac{R}{2}  \nabla_\nu \psi
- \frac{1}{2} H_{\mu\nu}\nabla^{\mu} (R \psi' \rho) - \frac{R \psi' \rho}{2} \vartheta u_\nu - H_{\nu\mu} \frac{R \psi' \rho}{2} u^{\rho}\nabla_{\rho} u^{\mu}\right)+
\end{equation*}
\begin{equation*}
- \ar \ell^2 \rho_{\fid}\left[ - g^{\alpha\beta}R^{\rho}_{\nu\alpha\mu}\nabla_{\beta} t^{\mu}_{\rho} +\nabla_{\mu} \left( R^{\rho}_{\sigma \rho \nu} t^{\sigma \mu} +R^{\mu}_{\sigma \rho \nu} t^{\sigma \rho} \right) \right. \end{equation*}
\begin{equation}
 \left.
\;\,\,\;\;+ R_{\sigma\nu} \nabla_{\mu}t^{\mu\sigma} + H_{\mu\nu} \nabla^{\mu} W +
 W \vartheta u_{\nu} + H_{\mu\nu}W u^{\rho}\nabla_{\rho} u^{\mu}
 \right].
\end{equation}

 Notice that this expression contains the full Riemann tensor, and in particular the Weyl tensor. Therefore,
 this kind of non-minimally coupled matter can have nontrivial behavior even in Ricci-flat spacetimes (as, for instance, Schwarzschild spacetime).

In the weak field limit, these will reduce to:
\begin{equation}
\nabla^{\mu}T_{\mu\nu}^{\std} = 0, 
\end{equation}
the one for the minimally coupled fluid, as expected. However, given that the fluid variables $\rho, p$ entering the definition of $T_{\mu\nu}^{\std}$ are not directly accessible, for the dark matter fluid these should be inferred by the sources of the gravitational field, \ie they should be written in terms of the components of $T^{\eff}_{\mu\nu}$. In terms of these components, the fluid equations are not of the standard form.

In the case of pressureless DM:
\begin{equation}
\rho_{\eff} = T^{\eff}_{\mu\nu}u^{\mu}u^{\nu} = \rho + \as \ell^2 \rho_{\fid} (\Box \psi + \partial_0^2 \psi)
- \frac{\ar}{2} \ell^2 \rho_{\fid}\left( +\Box \xi +\partial_0^2 \xi \right),
\end{equation}
\begin{equation}
3p_{\eff} = T^{\eff}_{\mu\nu}H^{\mu\nu} = 0+\as \ell^2 \rho_{fid}\left( -3 \Box \psi + H^{\mu\nu}\partial_{\mu}\partial_{\nu}\psi \right) - 3\frac{\ar}{2}\ell^2 \rho_{\fid} \partial_0^2 \xi.
\end{equation}
Constructing these and other quantities one has to re-express $\rho,p$ in terms of $\rho_{\eff},p_{\eff}$.

We will not discuss the equation of motion in full generality here, leaving this issue for further works. However we will sketch the effects of the non-minimal coupling in a particular case. Take for instance the case of scalar coupling ($\ar=0$), neglecting time derivatives:
\begin{equation}
\rho_{\eff} = \rho + a \rho_{\fid} \nabla^2 \psi,
\end{equation}
\begin{equation}
p_{\eff} = -\frac{2}{3}a \rho_{\fid} \nabla^2 \psi,
\end{equation}
whence
\begin{equation}
\rho = \rho_{\eff} - \frac{3}{2} p_{eff}.
\end{equation}

Notice that from the effective stress energy tensor we can also infer the effective EOS. In this simple case:
\begin{equation}
w \approx - \frac{2}{3}a \frac{\rho_{\fid}}{\rho} \nabla^2 \psi.
\end{equation}
Given that this is related to the speed of sound, the stability of the model requires $a \geq 0$ (and the Laplacian operator defined by a Euclidean metric is always negative definite).

A thourough stability analysis is required in more general cases.


\begin{thebibliography}{99}

\bibitem{WMAP}E. Komatsu et al.,
\emph{Seven-year Wilkinson Microwave Anisotropy Probe (WMAP) observations:cosmological interpretation}
ApJS $\bf{192}$ (2011)18 [arxiv:1001.4538v3 [astro-ph.CO]].

\bibitem{will}Clifford M. Will, 
\emph{The Confrontation between General Relativity and Experiment}, 
Living Rev. Relativity 9,  (2006),  3. URL: 
http://www.livingreviews.org/lrr-2006-3.

\bibitem{ferreira}
  P.~G.~Ferreira and G.~Starkmann,
  \emph{Einstein's Theory of Gravity and the Problem of Missing Mass,}
  Science {\bf 326} (2009) 812
  [arXiv:0911.1212 [astro-ph.CO]].

\bibitem{rot1}G. Gentile, P. Salucci, U. Klein, D. Vergani and P. Kalberla, 
\emph{The cored distribution of dark matter in spiral galaxies},
Mon. Not. Roy. Astron. Soc. $\bf{351}$(2004) 903 [arXiv:astro-ph/0403154v1].

\bibitem{rot2}G. Gentile, P. Salucci, U. Klein and G.L.Granato, 
\emph{NGC 3741:dark halo profile from the most extended rotation curves},
Mon. Not. Roy. Astron. Soc. $\bf{375}$(2007) 199 [arXiv:astro-ph/0611355v1].

\bibitem{cuco} W.J.G. De Blok,
\emph{The core-cusp problem},
Advances in Astronomy, vol. 2010, Article ID 789293, 14 pages, 2010[ arXiv:0910.3538v1[astro-ph.CO]].

\bibitem{TFR} R. B. Tully, J. R. Fisher,
\emph{A new method of determining distances to galaxies},
A\& A. $\bf{54}$, 661.

\bibitem{Don} F. Donato et al.,
\emph{A constant dark matter halo surface density in galaxies},
Mon. Not. R. Astron. Soc. $\bf{397}$, 1169-1176 (2009) [arXiv:0904.4054v1 [astro-ph.CO]].

\bibitem{Sal} G. Gentile, B. Famaey, H. Zhao and P. Salucci,
\emph{Universality of galactic surface densities within one dark halo scale-length},
Nature $\bf{461}$, 627-628 (2009) [arXiv:0909.5203v1 [astro-ph.CO]].

\bibitem{strig} L. E. Strigari, J. S. Bullock, M. Kaplinghat, J. D. Simon, M. Geha, B. Willman and M. G. Walker,
\emph{A common mass scale for satellites galaxies of the Milky Way},
Nature $\bf{454}$(2008)1096 [arXiv:0808.3772v1[astro-ph.CO]].

\bibitem{SN} R.S. de Souza, L. F. S. Rodrigues, E. E. O. Ishida and R. Opher,
\emph{The effect of a single supernova explosion on the cuspy density profile of a small-mass dark matter halo},
Mon. Not. R. Astron. Soc. $\bf{414}$, (2011) [arXiv:1104.2850v1[astro-ph.CO]].

\bibitem{momang} C.Tonini, A. Lapi and P. Salucci,
\emph{Angular momentum transfer in dark matter halos: ereasing the cusp},
Astrophys. J. {\bf 649} (2006) 591-598 [arXiv:astro-ph/0603051v3].

\bibitem{BTFR} S. Gurovich, S. S. McGaugh, K. C. Freeman, H.Jerjen, L. Staveley-Smith and W. J. G. De Blok
\emph{The baryonic Tully-Fisher relation},
Astrophys.J. $\bf{722}$ (2008) 1096-1097 [arXiv:astro-ph/0411521v1].

\bibitem{MOND}M. Milgrom,
\emph{A modification of the Newtonian dynamics as a possible alternative to the hidden mass hypothesis},
Astrophys. J. $\bf{270}$ (1983) 365-370.

\bibitem{alternative} J. D. Bekenstein,
\emph{Modified gravity as an alternative to dark matter},
Particle Dark Matter: Observations, Models and Searches', edited by G. Bertone (Cambridge U. Press, Cambridge 2010) Chap.6, p.95-114, [arXiv:1001.3876v1 [astro-ph.CO]].

\bibitem{MOND2}M. Milgrom,
\emph{MD or DM? Modified dynamics at low accelerations vs dark matter}
PoS HRMS2010 (2010) 033 .

\bibitem{STVGT} J. W. Moffat,
\emph{Scalar-tensor-vector gravity theory},
JCAP {\bf 03}(2006)004.

\bibitem{Bim} M. Milgrom,
\emph{Bimetric MOND gravity},
Phys. Rev. D {\bf 80}123536 (2009) [arXiv:0912.0790v2[gr-qc]].


\bibitem{fieldMOND} J-P, Bruneton, J. Esposito-Far\'ese,
\emph{Field-theoretical formulations of MOND-like gravity},
Phys. Rev. D $\bf{76}$, 124012 (2007) [arXiv:0705.4043v2 [gr-qc]].

\bibitem{MONDCMB}C. Skordis, D.F. Mota, P.G. Ferreira and C. Boehm, 
\emph{Large scale structures in Bekenstein's theory of relativistic modified Newtonian dynamics},
Phys. Rev. Lett. {\bf 96} (2006) 011301.

\bibitem{RMDM} J.P. Bruneton, S. Liberati, L. Sindoni and B. Famaey,
\emph{Reconciling MOND with Dark Matter?}
JCAP $\bf{03}$ (2009) 021[arXiv:0811.3143v1 [astro-ph]].

\bibitem{MONDark} 
Sanders, R.~H. \emph{Neutrinos as cluster dark matter}\ 2007, MNRAS, 
380, 331.

\bibitem{mondark2}
G.~W.~Angus, H.~Shan, H.~Zhao and B.~Famaey,
  \emph{On the Law of Gravity, the Mass of Neutrinos and the Proof of Dark
  Matter,}
  Astrophys.\ J.\  {\bf 654} (2007) L13
  [arXiv:astro-ph/0609125].

\bibitem{mondark3} Richtler, T., Schuberth, Y., Hilker, M., Dirsch, B., Bassino, L., \& Romanowsky A.~J.,
\emph{The dark matter halo of NGC 1399 - CDM or MOND?}
, \ 2008, Astronomy and Astrophysics, 478, L23.

\bibitem{khoury} K. Hinterbichler, J. Khoury and H. Nastase,
\emph{Towards a UV Completion for Chameleon Scalar Theories},
JHEP03(2011)061, Erratum-ibid JHEP06(2011)072 [arXiv:1012.4462v3[hep-th]]

\bibitem{chameleon}J. Khoury and A. Weltman,
\emph{Chameleon Fields: Awaiting Surprises for Tests of Gravity in Space},
Phys. Rev. Lett. {\bf 93}(2004)171104 [arXiv:astro-ph/0309300v3]

\bibitem{conlon} J. P. Conlon and F. G. Pedro,
\emph{Moduli-Induced Vacuum Destabilization},
JHEP(05(2011)079 [arXiv:1010.2665v1[hep-th]]

\bibitem{wett} J. Beyer, S. Nurmi and C. Wetterich,
\emph{Coupled dark energy and dark matter from dilaton anomaly},
Phys. Rev. D {\bf 84}(2011)023010 [arXiv:1012.1175v1[astro-ph.CO]]

\bibitem{goen} H. F. M. Goenner,
\emph{Theories of Gravitation with non-minimal Coupling of Matter and the Gravitational Field},
Found. Phys.{\bf 14}9(1984).

\bibitem{bertolami1}
  O.~Bertolami, J.~Paramos, T.~Harko and F.~S.~N.~Lobo,
  \emph{Non-minimal curvature-matter couplings in modified gravity,}
  [arXiv:0811.2876 [gr-qc]].

\bibitem{LmR}O. Bertolami, F. S. N. Lobo and J P\'aramos,
\emph{non-minimal coupling of perfect fluids to curvature},
Phys.\ Rev.\ D {\bf 78}, 2008.

\bibitem{bertolami2}
  O.~Bertolami and M.~C.~Sequeira,
  ``Energy Conditions and Stability in f(R) theories of gravity with
  non-minimal coupling to matter,''
  Phys.\ Rev.\  D {\bf 79}, 104010 (2009)
  [arXiv:0903.4540 [gr-qc]].

\bibitem{bertolami3}   O.~Bertolami and J.~Paramos,
  \emph{Mimicking dark matter through a non-minimal gravitational coupling with
  matter},
  JCAP {\bf 1003}, 009 (2010)
  [arXiv:0906.4757 [astro-ph.GA]].


\bibitem{weaklens}
  R.~Gavazzi {\it et al.},
  \emph{The Sloan Lens ACS Survey. 4. The mass density profile of early-type
  galaxies out to 100 effective radii,}
  Astrophys.\ J.\  {\bf 667} (2007) 176
  [arXiv:astro-ph/0701589].

\bibitem{weaklens2}Tian, L., Hoekstra, H., 
\& Zhao, H.\ 2009, MNRAS, 393, 885.

\bibitem{weaklens3} Pastor Mira, E., Hilbert, S., Hartlap, J., \& Schneider, P.\ 2011, astronomy and astrophysics, 531, A169.

\bibitem{far} V. Faraoni,
\emph{Cosmology in Scalar-Tensor Gravity},
Kluwer Academic Publishers, The Netherlands (2004).


\bibitem{harkorot}T. Harko, \emph{Galactic rotation curves in modified gravity with non-minimal coupling between matter and geometry},
Phys. rev. D {\bf 81}, 2010

\bibitem{harkoBEC}C. G. B\"{o}mer, T. Harko,
\emph{Can dark matter be a Bose-Einstein condensate?},
JCAP06(2007)025 [arXiv:0705.4158v4 [astro-ph]].

\bibitem{hu}W. Hu, R. Barkana and A. Gruzinov,
\emph{Fuzzy Cold Dark Matter: The wave Properties of Ultralight Particles},
Phys. Rev. Lett. {\bf 85} (2000)1158

\bibitem{chavanis}P. H. Chavanis,
\emph{Mass-radius relation of NEwtonian self-gravitating Bose-Einstein condensates with short-range interactions: I. Analytical results},
Phys. Rev. D {\bf 84}(2011) 043531 [arXiv:1103.2050v2 [astro-ph.CO]]

\bibitem{woo} T. P. Woo and T. Chiueh,
\emph{High-resolution simulation on structure formation with extremely light bosonic dark matter},
Astrophys. J. {\bf 697} (2009) 850

\bibitem{disf} J. D. Bekenstein, 
\emph{Relation between physical and gravitational geometry},
Phys Rev. D {\bf 48}, 1993.

\bibitem{adams}A. Adams, N. Arkani-Hamed, S. Dubovsky, A. Nicolis and R. Rattazzi,
\emph{Causality, analyticity and an IR onstruction to UV completion},
JHEP10(2006)014  [arXiv:hep-th/0602178v2]

\bibitem{mota}M. Zumalac\'arregui, T. S. Koivisto, D. F. Mota and P. Ruiz-Lapuente,
\emph{Disformal scalar fields and the dark sector of the universe},
JCAP05(2010)038

\bibitem{disfinfl} N. Kaloper,
\emph{Disformal inflation},
Phys. Lett. B, {\bf 583}(2004) 1-13

\bibitem{kunz} M. Kunz, A. R. Little, D. Parkinson and C. Gao,
\emph{Constraining the dark fluid},
Phys. Rev. D {\bf 80} (2009) 083533 [ arXiv:0908.3197v2 [astro-ph.CO]]

\bibitem{cota} A. Aviles, J. L. cervantes-Cota,
\emph{The dark degeneracy and interacting cosmic components},
arXiv:1108.24.57v1 [astro-ph.CO]

\bibitem{PFA} J. D. Brown,
\emph{Action functionals for relativistic perfect fluids},
Class. Quantum Grav. $\bf{10}$ (1993) 1579-1606.

\bibitem{caustics} A. Natarajan, P. Sikivie,
\emph{The inner caustics of dark matter halos},
Phys.Rev. D {\bf 73 }(2006) 023510 [arXiv:astro-ph/0510743v1].


\end{thebibliography}
\end{document}